\documentclass[aps,pre,amsmath,amssymb,reprint,showkeys,showpacs]{revtex4-1}

\newcommand{\nn}{\nonumber}

\newcommand{\ep}{\epsilon}

\newcommand{\be}{\begin{eqnarray}}
\newcommand{\ee}{\end{eqnarray}}

\newcommand{\bs}{\begin{subequations}\begin{eqnarray}}
\newcommand{\es}{\end{eqnarray}\end{subequations}}
\newcommand{\ba}{\begin{array}}
\newcommand{\ea}{\end{array}}
\newcommand{\bi}{\begin{itemize}}
\newcommand{\ei}{\end{itemize}}
\newcommand{\bei}{\begin{enumerate}}
\newcommand{\eei}{\end{enumerate}}

\usepackage{graphicx}
\usepackage{bm}

\usepackage{subfigure}
\begin{document} 


\title{DC electric field effect on the anomalous exponent of the hopping conduction in the one-dimensional  disorder model }

\author{Takeshi Egami}\email{egami.takeshi@canon.co.jp}
\author{Koshiro Suzuki}\email{suzuki.koshiro@canon.co.jp} 
\author{Katsuhiro Watanabe}\email{watanabe.katsuhiro@canon.co.jp} 
\affiliation{Analysis Technology Development Center, Canon Inc., 30-2
Shimomaruko 3-chome, Ohta-ku, Tokyo 146-8501, Japan}

\date{\today}
\begin{abstract}
DC electric field effect on the anomalous exponent of the hopping
 conduction in the disorder model is investigated. First, we explain the
 model and derive an analytical expression of the effective waiting time
 for the general case.  We show that the exponent depends on the
 external field.  Then we focus on a one-dimensional system in order to
 illustrate the features of the anomalous exponent. We derive
 approximate expressions of the anomalous exponent of the system
 analytically. For the case of a weak field, the anomalous exponent is
 consistent with that of diffusive systems. This is consistent with the
 treatments of Barkai et al. [Phys. Rev. E {\bf 63}, 046118 (2001)] and
 our result supports their theory.  On the other hand, for the case of a
 strong field and a strong disorder, the time evolution of the exponent
 clearly differs from that in the weak field.  The exponent is
 consistent with the well-known expression of the anomalous exponent in
 the Multiple Trapping Model at mesoscopic time scales. In the long time
 limit, a transition of the anomalous exponent to the same value of the
 weak field occurs. For the case of a strong field and a weak disorder,
 the exponent is equal to 1 and thus the diffusion is normal.  These
 findings are verified by the Monte Carlo simulation.
\end{abstract}
\pacs{05.40.Fb,05.60.-k,02.50.Ey}
\maketitle 

	\section{Introduction}\label{Introduction}
Hopping conductance is a dominant mechanism of electric conductivity in noncrystalline materials.
In the hopping conductance, each charge carrier hops from one localized state to another with the aid of thermal activation.
Moreover, it is widely known that the collective behavior of carriers exhibits anomalous diffusion (subdiffusion)  \cite{Xerox1}. 
Experimentally, one can observe the anomaly in the long-tail of the time-of-flight (TOF) signal  \cite{Xerox1, MP, Xerox2}.

Anomalous diffusion is characterized by the anomalous exponent $\alpha$, 
and has been studied in the context of continuous-time random walk (CTRW) thoroughly \cite{PhysRep.339.1}. 
This exponent exhibits itself in the power-law behaviors of the mean and mean-squared displacement with respect to time $t$. The diffusion is referred to as the subdiffusion if $0 < \alpha < 1$ holds.
In this case, the anomalous exponent and the exponent of 
 the waiting time of the	walker, 
$w(t) \sim t^{-(1+\alpha)}$, 
 are consistent with each other.  
As generalized cases, the accelerating and retarding anomalous
diffusions are known in various situations, both theoretically and
experimentally \cite{JPhysA.45.145001,PhysRevE.66.046129,SCK2004,
PhysRevE.78.021111, PhysRevLett.54.616,PhysRevLett.103.018102,
PhysRevE.71.041915, Weiss20043518}.  In these cases, {\it $\alpha$
depends on time, i.e., $\alpha=\alpha(t)$}. To be precise, $\alpha(t)$
increases (decreases) as the time evolves in the accelerating
(retarding) anomalous diffusion.
%

%
It is well known that CTRW has succeeded in describing the long-tail of
the TOF signals in a certain range of time scales by treating $\alpha$
as a fitting parameter \cite{Xerox1}.  However, the relation of $\alpha$
to physical quantities of interest, such as the external field, the
density of state (DOS) of the trap levels, or the spatial structure of
the hopping sites, cannot be addressed in the framework of CTRW. Hence,
it is significant to study this issue for physically understanding the
mechanism of the hopping conductance.

Such relations have been partially given by two representative models of
the hopping conductance, i.e., the "Multiple Trapping Model (MTM)"
\cite{SolidStateCommun.37.49,TMMA,MTM}, and the "Scher-Montroll model
(SMM)" \cite{Xerox1,Xerox2,SL1,SL2}.  The MTM is a well known
phenomenological model. In this model, carriers are trapped for certain
times which are determined from the depths of trap levels, then released
and trapped again in an another one.  Repeating this process, the
distribution of the carriers in the trap levels in the long time limit
is obtained. Using this, the mobility is determined. It is worth
emphasizing that {\it the drift process of carriers is treated as a
single trap level problem} \cite{SolidStateCommun.37.49}.
The anomalous exponent is related to the typical width of the DOS of the
trap levels $T_c$ and the temperature $T$ as
$
\alpha = T/T_c
$
\cite{SolidStateCommun.37.49}.
The DOS is assumed to be of the exponential type,
$p(\varepsilon) = e^{(\varepsilon-\varepsilon_C)/(k_B T_c)}/(k_B T_c)$ ($\varepsilon\leq \varepsilon_C$),
where $\varepsilon$ is the energy and $\varepsilon_C $ is that of the edge of the conduction band.
This type of DOS is typical for disordered irrorganic semiconductors and well explains the experimental results   \cite{TMMA}. 
Although $\alpha$ is expressed in such a simple form, the relation of
$\alpha$ to the external field and the spatial structure of hopping
sites is unclear, since they are not considered in the MTM.
%

The SMM is an effective media model in mesoscopic scales.
In the SMM, spatial structures of hopping sites are partially taken into
account. 
The number density of the hopping sites is assumed to be constant, the
lattice of the sites are smeared out into a continuum media, and the
hopping activation energy is replaced by its average value
\cite{SL1,SL2,Xerox1}.
The exact form of the waiting time is obtained with the aid of the mathematical technique in  reference \cite{CCSB}. 
However, the relation of $\alpha$ to the external field, the DOS and the
microscopic spatial structure of hopping sites is unclear, since they
are not taken into account in the SMM.

On the other hand, two microscopic models have been used in order to
study the hopping conductance. 
One is the "polaron model"  \cite{FKBN01,AnnPhys.8.343,SMT}, and  the other is the "disorder model"  \cite{Bassler}, which is studied in this paper. In these models, the external field, the microscopic spatial structure of the hopping sites, and the DOS are taken into account.
The polaron model is suitable for a system with strong electron-phonon couplings, and with relatively negligible effects of the	energy disorder  \cite{AEKBNWB}.
In the polaron model, 
the strong electron-phonon couplings strongly distort the surroundings of a carrier which is in the localized
 state,
 and this distortion lowers the energy of the carrier (self-trapping).
Because the carrier moves together with the associated distortion, the carrier with the distortion can be regarded as a quasi-particle which is referred to as the polaron. 
As long as the authors know, the relation of $\alpha$ of the polaron model to the physical quantities is still an open issue.

In contrast to the polaron model, the disorder model is suitable for a system where the anomalous charge transport is dominantly activated by the static energy disorder of the hopping sites,	and where the effect of the weak electron-phonon coupling is relatively negligible. 
In the disorder model  \cite{Bassler}, as we will briefly review later, the hopping rate between two sites is described by the conventional model of  \cite{AHL}, which is based on  \cite{MA}. 
To solve the model of  \cite{AHL}, a Monte Carlo (MC) simulation has been performed 
 \cite{Bassler}, and the effect of the electric field to the mobility has been studied recently \cite{PhysRevB.86.045207}. 
From the viewpoint of the anomalous exponent, 
we have shown the relation of $\alpha$ to the microscopic spatial structure of the hopping sites for a diffusive system  \cite{ESW}. 
In addition, we have also shown that $\alpha$ depends on time.
However, the effect of the external field on $\alpha$ is still unknown. 
In this paper, we study this issue.

The paper is organized as follows.
In section \ref{Theory}, we present our theory.
In particular, we explain the model we study and its related
issues in section \ref{Model}, and we calculate the exact form of the effective waiting time and derive the asymptotic form of the 
anomalous exponent theoretically in section \ref{Alpha}. 
In section \ref{Simulation}, the result is verified by the MC simulation. 
In section \ref{Discussion}, we discuss the relation of our results to the previous studies.
In section \ref{Summary}, we summarize our study.

		\section{Theory}\label{Theory}

\subsection{Model}\label{Model}
In this study, we consider the disorder model  \cite{Bassler}.  
{
The basic framework is essentially the same as the one in Ref.~\cite{ESW}, except that an external field acts on the carriers. 
}
In the following, we explain the DOS, the hopping rate and the waiting
time of the model which we use in this study.

\subsubsection{Density of states}

\begin{figure}[tbh]
	\subfigure[Overall view]
	{
		\includegraphics[width=4.0cm]{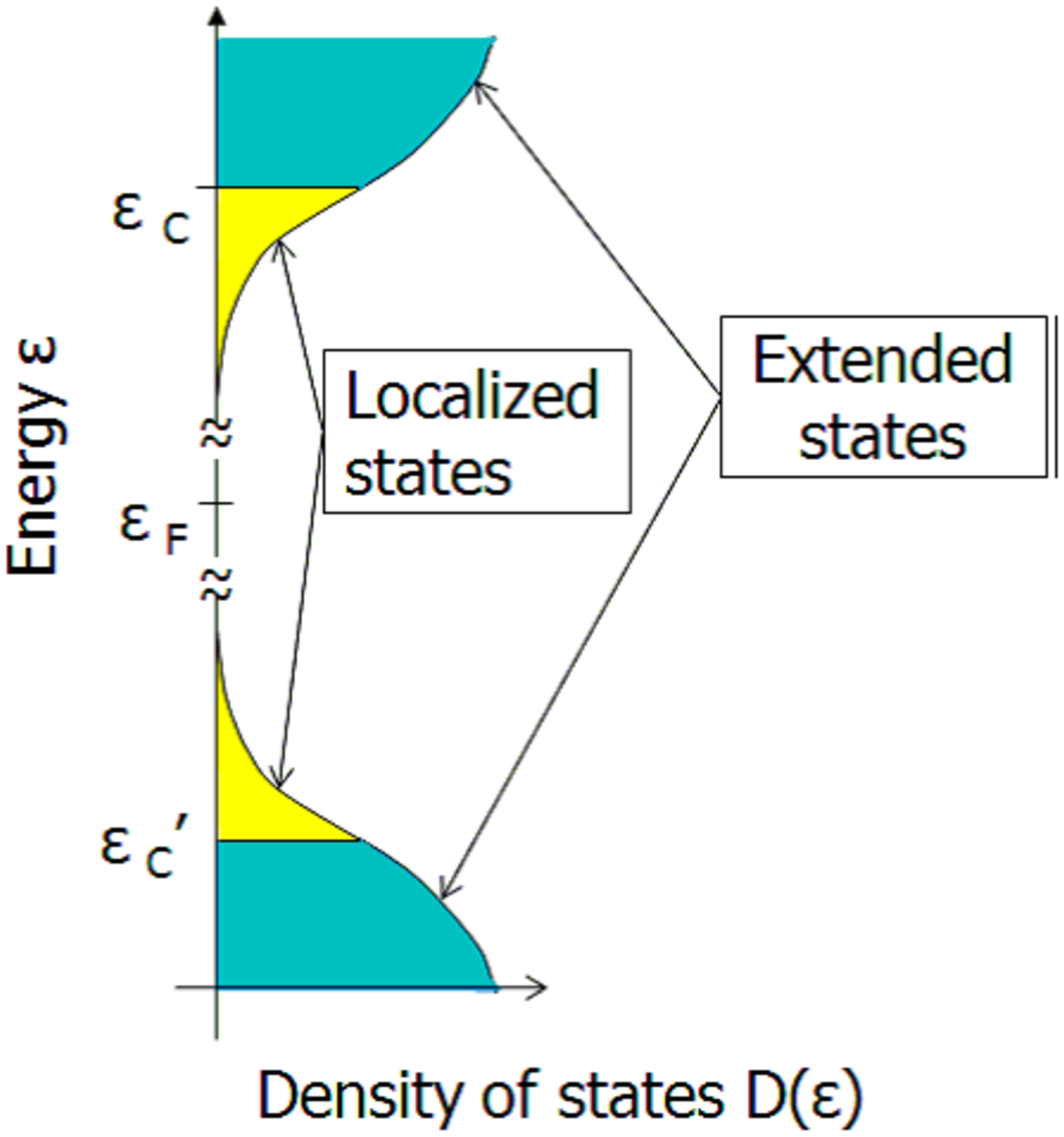}
		\label{DOS1}
	}
	\subfigure[Exponential tail]
	{
		\includegraphics[width=4.0cm]{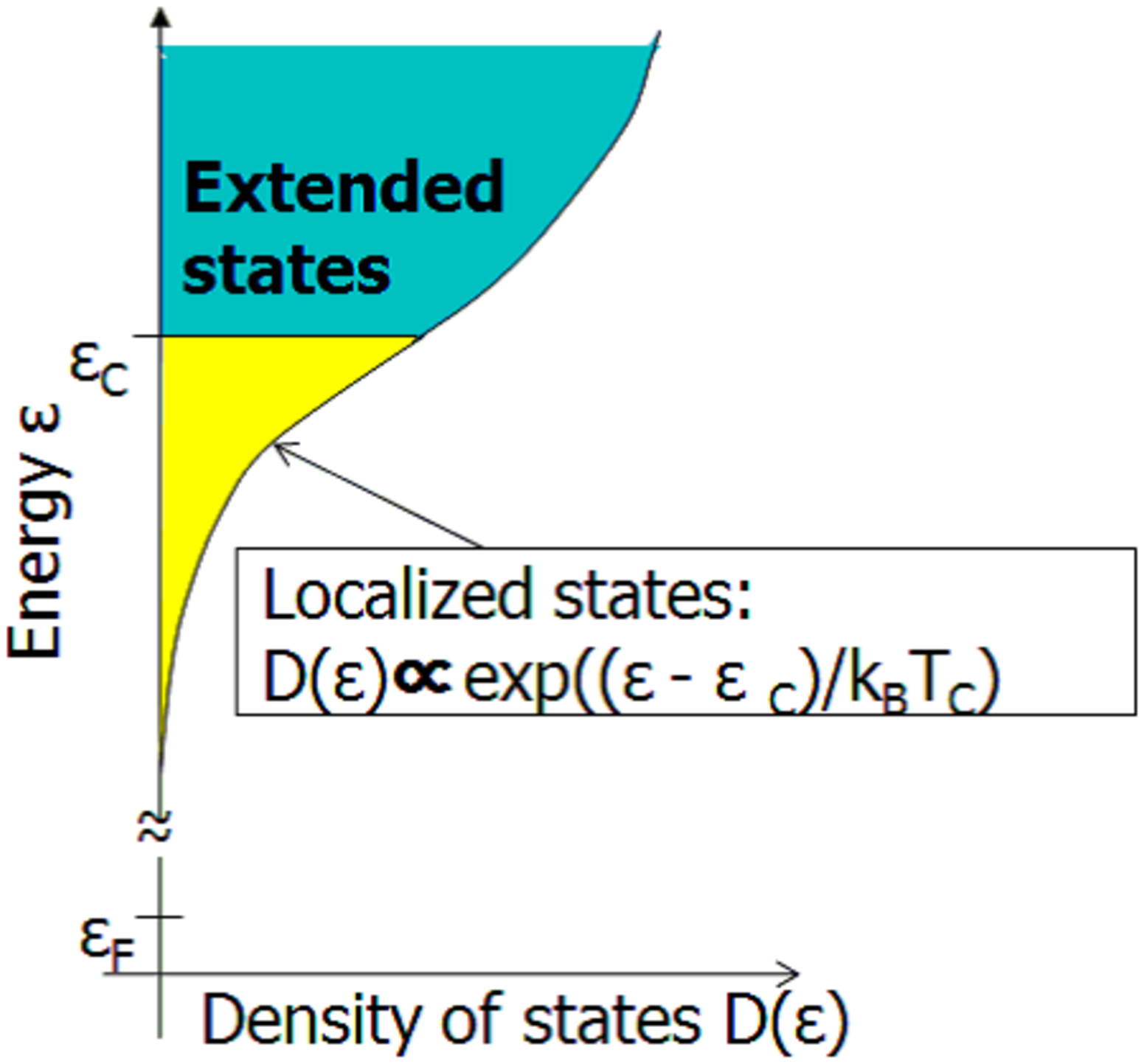}
		\label{DOS2}
	}%
		\caption{Schematic figures of DOS: (a) Overall view: The
 conduction and valence bands (extended states) and their tails (localized states). States in the band tails are separated from the conduction (valence) bands 	by the mobility edges $\epsilon_C$ ($\epsilon_C'$).
	 \ \  (b): Exponential tail of the conduction band. Its typical
 width is denoted as $T_c$.}
\end{figure}

Amorphous semiconductors have conduction and valence band tails which
are shown schematically in Fig.~\ref{DOS1}. The existence of the tail
is first pointed out by Urbach  \cite{PhysRev.92.1324} and it is referred to as the Urbach tail. 
The states in the band tails are localized; they are separated from the extended states by the critical energy which is referred to as the mobility edge \cite{JPhysC.2.1230}. 
The energies $\varepsilon_C$ and $\varepsilon_C'$ in Fig.~\ref{DOS1}
are the mobility edges in the conduction and the valence bands, respectively. 
Tiedje et al. have shown that the exponential band tail well explains
the experimental results of the temperature dependence of the anomalous
exponent for irrorganic amorphous semiconductors  \cite{TMMA}. 
A recent calculation of the electronic density of states also supports that the band tails are well approximated by the exponential functions
 \cite{PhysRevB.83.045201}.
In this study, we adopt as the DOS the exponential type. 
Then the DOS can be expressed as 
\begin{align}
p(\varepsilon) &= \frac{1}{k_B T_c}e^{(\varepsilon-\varepsilon_C)/(k_B T_c)},
\label{13040301}
\end{align}
where $\varepsilon\leq \varepsilon_C$  and $T_c$ is the typical width of the DOS (see Fig.~\ref{DOS2}).
{
}

\subsubsection{Hopping rate between two sites}

The intuitive picture of the hopping conduction is that carriers hop between the localized states. 
This implies that the hopping conduction is dominant at low temperatures where thermal activation of the carriers to the extended states, which results in the band conduction, can be neglected.  
In such situations, we can adopt the approximation that
the states above (below) the Fermi energy $\varepsilon_F$ are unoccupied (occupied). 
Then, the carriers hop between the empty states in the conduction band
tail, whose DOS is approximated by Eq.~(\ref{13040301}) (localized states in Fig.~\ref{DOS2}).

In this study, as an electric field $\vec{E}(\vec{r})$ at the position
$\vec{r}=(x,y,z)$, we consider a constant field $\vec{E}=(-E, 0, 0)$
($E\geq 0$) in the negative direction of the $x$ axis. Under this
electric field, an external force $\vec{F}=(F, 0, 0)\equiv(eE,0,0)$ acts
on each carrier with charge $-e$. Here, $e (>0)$ is the elementary
charge.  In addition, assuming that the number of the carriers is small
enough compared to the number of the states, the occupation of the
 states can be neglected.
{
Then, the hopping probability per unit time of the carrier to hop from site $i$ to site $j$, which we denote $\nu_{ij}$, is approximately given as  \cite{AHL}
}
\begin{align}
\nu_{ij}
&=
\nu_0 e^{-2R_{ij}/\xi -(\epsilon_{ji}-Fx_{ji})\Theta(\epsilon_{ji}-Fx_{ji})/k_BT}.
\label{MA}
\end{align}
{
Here, $\xi$ is the localization length of the localized state, $\epsilon_i$ is the energy of site $i$ on the position $\vec{r}_i\equiv(x_i,y_i,z_i)$, $R_{ij}$ is the distance of sites $i$ and $j$, $x_{ji}\equiv x_j-x_i$ is the relative coordinate of sites $i$ and $j$ in the $x$ direction,  $\ep_{ji}\equiv \epsilon_j-\epsilon_i$ is the difference of the energy of site
$i$ and $j$, $\nu_0$ is the magnitude of the hopping rate, $\Theta(x)$
is the Heaviside's step function, where $\Theta(x)=1$ for $x>0$ and 0
otherwise, and $T$ is the temperature. 
}
This hopping rate is usually used in the disorder model and is well verified \cite{Bassler}.

\subsubsection{Waiting time }

In this sub-subsection, we briefly review the waiting time $w(t)$ of the CTRW
 \cite{PhysRep.339.1}, and that of the disorder model.  

{
In CTRW  \cite{PhysRep.339.1}, if the carrier is at the position $x=0$ at $t=0$, the probability distribution that the carrier is found at position $x$ at time $t$ is given by 
}
\begin{align}
\rho(x,t)
&=
\sum_{x' \neq x} \int_0^t d \tau \rho(x',\tau) \psi(x,x',t-\tau) + \Phi(x,t) \delta_{x0}.
\end{align}
{
Here, $\Phi(x,t) \equiv 1 - \int_0^t d\tau w(x,\tau)$, and 
$\psi(x',x,t)$ denotes the probability density that the carrier hops from $x$ to $x'$ after waiting time $t$.
Using $\psi(x',x,t)$, the waiting-time probability distribution $w(x,t)$ is expressed as
}
\begin{align}
w(x,t)
&=
\sum_{x' \neq x} \psi(x',x,t).
\nn
\end{align}
{
In the disorder model, it is assumed that $\psi(x',x,t)$ is given by the product of $w(t) $ and the spatial part of the probability distribution $\phi(x',x)$, i.e., $\psi(x',x,t) = w(t) \phi(x',x)$.
Here, $\phi(x',x)$ is normalized as $\sum_{x'} \phi(x',x) = 1$. 
In this study, we are interested in the time-dependent part $w(t)$.
}

{
Defining $\Lambda_i$ by $\Lambda_i \equiv \sum_{j\in\mathcal{N}} \nu_{ij}$ ($\mathcal{N}$ represents the set of all the sites in the system), we can express the waiting time of the carrier on site $i$ as
\begin{eqnarray}
w_i(t) = \Lambda_i e^{-\Lambda_i t}.
\label{wt}
\end{eqnarray}
The spatial distribution of the hopping sites causes the disorder of energy. 
In the disorder model, the disorder of the hopping sites is assumed to be moderate and their positions fluctuate around the structured lattice points. However, we incorporate the effect of the spatial disorder to that of the disorder of the site energy for the sake of simplicity, and an assumption that the sites are situated on structured lattice points with lattice spacing $a$ is used in this study. 
}

\subsection{Anomalous exponent of disorder model in the DC electric field}\label{Alpha}
\subsubsection{General system}
{
Now we calculate the anomalous exponent. 
We formulate general exact results for arbitrary spatial dimensional systems, where the carrier
at each site can hop to all the sites in the system. 
It is important that the waiting-time probability distribution,
Eq.~(\ref{wt}), depends on the energy of the site.
If the energy is distributed according to some probability distribution,
then the effective waiting time of the system, which we denote $\langle
w(t) \rangle_F$, is given by the ensemble average with respect to the
energy distribution \cite{Kivelson}. Its explicit expression is }
\begin{widetext}
\begin{align}
\langle w(t) \rangle_F&=
	\left(\prod_{j\in \mathcal{N}}\int_{-\infty}^\infty d \epsilon_{ji}
	p_L(\epsilon_{ij})
	\right)
	w_i(t)
\nn
\\&=
	\left(\prod_{j\in \mathcal{N}}\int_{-\infty}^\infty\!\! d \epsilon_{ji}
	\frac{e^{-\frac{|\ep_{ji}|}{k_BT_c}}}{2k_BT_c}
	\right)
	\left(
			\sum_{k\in\mathcal{N}}\!\! K_{ki}
			e^{
				-\frac{(\epsilon_{ki}-Fx_{ki})\Theta(\epsilon_{ki}-Fx_{ki})}{k_BT}}
			e^{
				-\sum_{l\in\mathcal{N}} K_{li}
			e^{-\frac{(\epsilon_{li}-Fx_{li})\Theta(\epsilon_{li}-Fx_{li})}{k_BT}} t
			}
	\right),
	\label{wtev}
\end{align}
\end{widetext}
where
\begin{align}
K_{ji}&\equiv \nu_0e^{-2R_{ji}/\xi }.
\label{defK}
\end{align}
Here, $\langle\dots\rangle_F$ denotes an ensemble average when the
carriers are subject to an external driving field.
{
We have omitted the subscript $i$ in the expression $\langle w(t) \rangle_F$, because it is assumed that 
the system has a translational invariance, 
as a result of the spatial coarse graining due to the integration with respect to site energies.  
In Eq.~(\ref{wtev}), we have used the fact that the energy difference $\ep_{ji}$ obeys the Laplace distribution $p_L(\ep_{ji}) \equiv e^{-|\ep_{ji}|/k_BT_c}/(2k_BT_c)$.
}
Using the transformations of integral variables  \cite{US}, we can calculate the integrals in Eq.~(\ref{wtev}) analytically. 
After tedious but straightforward calculations which we show in Appendix \ref{Appendix}, we obtain the effective waiting time as follows:
\begin{widetext}
\begin{align}
\langle w(t) \rangle_F
&=
	\sum_{k\in \mathcal{N}_+}
	\left\{
	K_{ki}I^{(1)+}_{ki}
	\prod_{j\in \mathcal{N}_+,j\neq k}I^{(0)+}_{ji}
	\prod_{m\in \mathcal{N}_0,m\neq k}I^{(0)0}_{mi}
	\prod_{l\in \mathcal{N}_-,l\neq k}I^{(0)-}_{li}
\right\}
\nn\\
&
+
	\sum_{k\in \mathcal{N}_0}
	\left\{
	K_{ki}I^{(1)0}_{ki}	
	\prod_{j\in \mathcal{N}_+,j\neq k}I^{(0)+}_{ji}
	\prod_{m\in \mathcal{N}_0,m\neq k}I^{(0)0}_{mi}
	\prod_{l\in \mathcal{N}_-,l\neq k}I^{(0)-}_{li}
	\right\}
\nn	
\\
&+
	\sum_{k\in \mathcal{N}_-}
	\left\{
	K_{ki}I^{(1)-}_{ki}	
	\prod_{j\in \mathcal{N}_+,j\neq k}I^{(0)+}_{ji}
	\prod_{m\in \mathcal{N}_0,m\neq k}I^{(0)0}_{mi}
	\prod_{l\in \mathcal{N}_-,l\neq k}I^{(0)-}_{li}
	\right\}.
	\label{11060902}
\end{align}
\end{widetext}
Here, $\mathcal{N}_+$ ($\mathcal{N}_-$) represents a set of neighboring
sites which satisfy the condition $x_{li}>0$ ($x_{li}<0$),
$\mathcal{N}_0$ represents a set of neighboring sites which satisfy the
condition $x_{li}=0$, and
\begin{widetext}
\begin{align}
I^{(0)+}_{li}	&\equiv
		\left(
		1
		-
		\frac{1}{2}e^{-\frac{Fx_{li}}{k_BT}\frac{T}{T_c}}
		\right)
		e^{
			- K_{li}t
		}	
	+
	\frac{1}{2}e^{-\frac{Fx_{li}}{k_BT}\frac{T}{T_c}}
		\frac{T}{T_c}(K_{li}t)^{-\frac{T}{T_c}}
		\gamma\left(\frac{T}{T_c},K_{li}t\right)	,
\nn\\
	I^{(1)+}_{li}	
	&\equiv
		\left(
		1
		-
		\frac{1}{2}e^{-\frac{Fx_{li}}{k_BT}\frac{T}{T_c}}
		\right)
		e^{
			- K_{li}t
		}	
	+
	\frac{1}{2}e^{-\frac{Fx_{li}}{k_BT}\frac{T}{T_c}}
		\frac{T}{T_c}(K_{li}t)^{-1-\frac{T}{T_c}}
		\gamma\left(\frac{T}{T_c}+1,K_{li}t\right),
	\nn\\
	I^{(0)0}_{li}		
	&\equiv
		\frac{1}{2}
	e^{
			-K_{li}t
		}			
	+
		\frac{1}{2}\frac{T}{T_c}
		(K_{li}t)^{-\frac{T}{T_c}}
		\gamma\left(\frac{T}{T_c},K_{li}t\right)	,
	\nn\\	
	 I^{(1)0}_{li}&\equiv
		\frac{1}{2}
	e^{
			-K_{li}t
		}	
	+
		\frac{1}{2}
		(K_{li}t)^{-1-\frac{T}{T_c}}
		\gamma\left(\frac{T}{T_c}+1,K_{li}t\right),
	\nn\\
	I^{(0)-}_{li}&\equiv
			\frac{1}{2}e^{\frac{Fx_{li}}{k_BT}\frac{T}{T_c}\left(1-\frac{T}{T_c}\right)}
			e^{-K_{li}e^{\frac{Fx_{li}}{k_BT}\frac{T}{T_c}}t}	
			+
			\frac{1}{2}e^{\frac{Fx_{li}}{k_BT}\frac{T}{T_c}}
		(K_{li}t)^{\frac{T}{T_c}}\gamma\left(1-\frac{T}{T_c},K_{li}t\right)
		\nn\\
		&
			-
			\frac{1}{2}e^{\frac{Fx_{li}}{k_BT}\frac{T}{T_c}}
		(K_{li}t)^{\frac{T}{T_c}}\gamma\left(1-\frac{T}{T_c},K_{li}e^{\frac{Fx_{li}}{k_BT}\frac{T}{T_c}}t\right)
	+
		\frac{1}{2}e^{-\frac{Fx_{li}}{k_BT}\frac{T}{T_c}}
		\frac{T}{T_c}(K_{li}t)^{-\frac{T}{T_c}}
		\gamma\left(\frac{T}{T_c},K_{li}e^{\frac{Fx_{li}}{k_BT}}t\right),
	\nn\\
	I^{(1)-}_{li}
	&\equiv
		\frac{1}{2}e^{Fx_{li}/k_BT_c}
	e^{
			-K_{li}t
		}
+
		\frac{1}{2}e^{\frac{Fx_{li}}{k_BT}\frac{T}{T_c}}
		\frac{T}{T_c}(K_{li}t)^{-1+\frac{T}{T_c}}
		\left[
		\gamma\left(1-\frac{T}{T_c},K_{li}t\right)
			-
			\gamma\left(1-\frac{T}{T_c},K_{li}e^{\frac{Fx_{li}}{k_BT}\frac{T}{T_c}}t\right)
		\right]
	\nn\\
	&	
	+
		\frac{1}{2}e^{-\frac{Fx_{li}}{k_BT}\frac{T}{T_c}}
		\frac{T}{T_c}(K_{li}t)^{-1-\frac{T}{T_c}}
		\gamma\left(\frac{T}{T_c}+1,K_{li}e^{\frac{Fx_{li}}{k_BT}}t\right).\nn
\end{align}
\end{widetext}
Here,
\be
	\gamma(T/T_c,K_{ij}t)\equiv \int_0^{K_{ij}t}d \tau
		\tau^{-1+T/T_c}e^{-\tau}
	\label{incompgamma}
\ee
 is the lower incomplete gamma function. 
Eq.~(\ref{11060902}) is the exact form of the effective waiting time of the disorder model with a constant external field.

Since the incomplete gamma function $\gamma\left(T/T_c,C_\gamma\right)$ coincides with the gamma function $\Gamma\left(T/T_c\right)$ at the same order of  $C_\gamma$ for the case $0.05\le T/T_c\le 1$
\cite{ESW}, one can see that there are three characteristic time scales
for  $|x_{ki}|$, i.e.,
\begin{align}
	\tau_{li1}\equiv K_{li}^{-1},\ \tau_{li2}\equiv K_{li}^{-1}e^{\frac{F|x_{li}|}{k_BT}\frac{T}{T_c}},\ \tau_{li3}\equiv K_{li}^{-1}
	e^{\frac{F|x_{li}|}{k_BT}}.
	\label{scale}
\end{align}
From the viewpoint of the dependence of the anomalous exponent on the
external field and disorder, it is important that $\tau_{li1}$ does not
depend on either of them, $\tau_{li2}$ depends on both of them, and
$\tau_{li3}$ depends only on $Fa/k_BT$.  It is clear from
Eq.~(\ref{scale}) that these time scales reduce to a single value
$\tau_{li1}=\tau_{li2}=\tau_{li3}=K_{li}^{-1}$ when no external field is
applied.  In other words, $\tau_{li2}$ and $\tau_{li3}$ are generated by
the external field.  We will see how these time scales characterize the
time evolution of the exponent in a one-dimensional system in section
\ref{Sec:1D}.

If no external field is applied to carriers, Eq.~(\ref{11060902}) yields
\begin{widetext}
\begin{align}
\langle w(t) \rangle_{0}
&=
	\sum_{k\in \mathcal{N}}
	\frac{1}{2}K_{ki}
	\left[	
			e^{
				- K_{ki}t
			}	
		+
			\frac{T}{T_c}(K_{ki}t)^{-1-\frac{T}{T_c}}
			\gamma\left(\frac{T}{T_c}+1,K_{ki}t\right)		
	\right]
\prod_{j\in \mathcal{N},j\neq k}
\left[
		\frac{1}{2}
		e^{
			- K_{ji}t
		}	
	+
	\frac{1}{2}
		\frac{T}{T_c}(K_{ji}t)^{-\frac{T}{T_c}}
		\gamma\left(\frac{T}{T_c},K_{ji}t\right)	
\right]
.
\label{12110501}
\end{align}
\end{widetext}
Here, $\langle\dots\rangle_0$ denotes an ensemble average when no external field is applied to the carriers. 
Eq.~(\ref{12110501}) is just the effective waiting time for diffusive systems which we derived in Ref.~\cite{ESW}.

\subsubsection{One-dimensional system}
\label{Sec:1D}
\begin{figure}[htb]
\includegraphics[width=8.5cm]{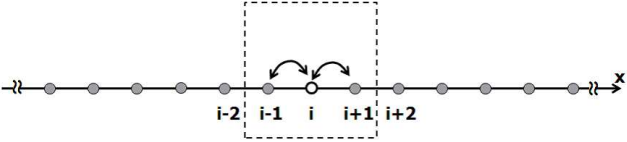}	
\caption{A schematic figure of the system which is considered. It is one-dimensional and hopping up to the first-nearest neighbors is allowed. 
}
\label{system}
\end{figure}
Although Eq.~(\ref{11060902}) is exact, its physical meaning and
dependence on the model parameters are not clear. 
In the following, in order to illuminate the features of the effective
waiting time, Eq.~(\ref{11060902}), we consider a one-dimensional system
where carriers can hop only to the nearest neighbors (Fig.~\ref{system})
and derive approximate expressions of the anomalous exponent by
extracting the dominant contribution from the neighboring sites and by
simplifying its analytic structure.  We concentrate on the time scale
where effects of the next nearest neighbors are negligible
($t<\tau_{2nd}\equiv\nu_0^{-1}e^{4a/\xi}$).
 
For the case mentioned above, Eq.~(\ref{11060902}) becomes
\begin{widetext}
\begin{align}
\langle w(t) \rangle_F
&=
	K_{1}
	\left[	
			\left(
			1
			-
			\frac{1}{2}e^{-\frac{Fa}{k_BT}\frac{T}{T_c}}
			\right)
			e^{
				- K_{1}t
			}	
		+
		\frac{1}{2}e^{-\frac{Fa}{k_BT}\frac{T}{T_c}}
			\frac{T}{T_c}(K_{1}t)^{-1-\frac{T}{T_c}}
			\gamma\left(\frac{T}{T_c}+1,K_{1}t\right)		
	\right]
\nn\\
&
\times
\left[
\frac{1}{2}e^{\frac{-Fa}{k_BT}\frac{T}{T_c}\left(1-\frac{T}{T_c}\right)}
			e^{-K_{1}e^{\frac{-Fa}{k_BT}\frac{T}{T_c}}t}	
			+
			\frac{1}{2}e^{\frac{-Fa}{k_BT}\frac{T}{T_c}}
		(K_{1}t)^{\frac{T}{T_c}}\gamma\left(1-\frac{T}{T_c},K_{1}t\right)
\right.
		\nn\\
		&
\left.
			-
			\frac{1}{2}e^{\frac{-Fa}{k_BT}\frac{T}{T_c}}
		(K_{1}t)^{\frac{T}{T_c}}\gamma\left(1-\frac{T}{T_c},K_{1}e^{\frac{-Fa}{k_BT}\frac{T}{T_c}}t\right)
	+
		\frac{1}{2}e^{\frac{Fa}{k_BT}\frac{T}{T_c}}
		\frac{T}{T_c}(K_{1}t)^{-\frac{T}{T_c}}
		\gamma\left(\frac{T}{T_c},K_{1}e^{\frac{-Fa}{k_BT}}t\right)
\right]
\nn\\
&+
	K_{1}
\left[
		\frac{1}{2}e^{\frac{-Fa}{k_BT}\frac{T}{T_c}}
	e^{
			-K_{1}t
		}
+
		\frac{1}{2}e^{\frac{-Fa}{k_BT}\frac{T}{T_c}}
		\frac{T}{T_c}(K_{1}t)^{-1+\frac{T}{T_c}}
		\left\{
		\gamma\left(1-\frac{T}{T_c},K_{1}t\right)
			-
			\gamma\left(1-\frac{T}{T_c},K_{1}e^{\frac{-Fa}{k_BT}\frac{T}{T_c}}t\right)
		\right\}
\right.
	\nn\\
	&
\left.
	+
		\frac{1}{2}e^{\frac{Fa}{k_BT}\frac{T}{T_c}}
		\frac{T}{T_c}(K_{1}t)^{-1-\frac{T}{T_c}}
		\gamma\left(\frac{T}{T_c}+1,K_{1}e^{\frac{-Fa}{k_BT}}t\right)	
\right]
\nn
\\
&
\times
\left[
		\left(
		1
		-
		\frac{1}{2}e^{-\frac{Fa}{k_BT}\frac{T}{T_c}}
		\right)
		e^{
			- K_{1}t
		}	
	+
	\frac{1}{2}e^{-\frac{Fa}{k_BT}\frac{T}{T_c}}
		\frac{T}{T_c}(K_{1}t)^{-\frac{T}{T_c}}
		\gamma\left(\frac{T}{T_c},K_{1}t\right)	
\right].
	\label{12110601}
\end{align}
\end{widetext}
Here, $K_1\equiv \nu_0e^{-2a/\xi }$.  
The three characteristic time scales in this system are the followings:
\begin{align}
	\tau_1\equiv K_{1}^{-1},\ \tau_2\equiv K_{1}^{-1}e^{\frac{Fa}{k_BT}\frac{T}{T_c}},\ \tau_3\equiv K_{1}^{-1}
	e^{\frac{Fa}{k_BT}}.
	\label{12121101}
\end{align}
 

Before further discussion, it is important to denote some relations we use later.
First, for the sites which satisfy the condition $K_{1}t\ll1$, the following relations hold \cite{ESW}:
\begin{align}
\frac{T}{T_c}\frac{\gamma\left(\frac{T}{T_c},K_{1}t\right)}{(K_{1}t)^{T/T_c}}
\simeq 1, \hspace{1em} & e^{-K_{1}t}\simeq 1.
\label{11071502}
\end{align}
Second, for the sites which satisfy the condition $K_{1}t \gg 1$, the following relations hold \cite{ESW}:
\begin{align}
\gamma(T/T_c, K_{1}t) \simeq \Gamma(T/T_c), \hspace{1em} &e^{-K_{1}t}\simeq 0.
\label{approx}
\end{align}
Here, $\Gamma(T/T_c)$ is the Gamma function.
Now we derive approximate expressions of the anomalous exponent,
Eq.~(\ref{12110601}), for the cases of weak and strong fields,
respectively.
\ \vspace{1em}\\
{\bf Weak field cases:}
First, we consider the weak field case where  the condition $\tau_1\sim \tau_2\sim \tau_3$ holds. 
In other words, we consider the case where $e^{Fa/k_BT}\simeq 1+Fa/k_BT$ holds.

Substituting $e^{Fa/k_BT}\simeq 1+Fa/k_BT$ into Eq.~(\ref{12110601}),  we obtain 
\begin{align}
		\langle w(t) \rangle_{\mathrm{w}}
	&\simeq
K_1	
	\left[	e^{
				- K_1t
			}	
		+
			\frac{T}{T_c}(K_{1}t)^{-1-\frac{T}{T_c}}
			\gamma\left(\frac{T}{T_c}+1,K_{1}t\right)		
	\right]
\nn\\
&
\times
\frac{1}{2}	
\left[
		e^{
			- K_{1}t
		}	
	+
		\frac{T}{T_c}(K_{1}t)^{-\frac{T}{T_c}}
		\gamma\left(\frac{T}{T_c},K_{1}t\right)	
\right]
\nn
\\
&+\mathcal{O}\left(\frac{Fa}{k_BT}\right).
\label{12110801}
\end{align}
In order to obtain an asymptotic analytic form of the anomalous exponent, we consider a time scale where $t\gg \tau_1$ holds. 
From Eq.~(\ref{approx}), we obtain the following approximate expression
for Eq.~(\ref{12110801}),
\begin{align}
		\langle w(t) \rangle_{\mathrm{w}}
	&\simeq
	K_1
	\frac{1}{2}
			\left(\frac{T}{T_c}\right)^2
			(K_{1}t)^{-1-2\frac{T}{T_c}}
			\Gamma\left(\frac{T}{T_c}+1\right)		
		\Gamma\left(\frac{T}{T_c}\right)	
	\nn
	\\
	&+\mathcal{O}\left(\left(\frac{Fa}{k_BT}\right)^2\right)	
		\propto t^{-1-2\frac{T}{T_c}}.
\label{12110802}
\end{align}
It is worth noting that the leading correction term is quadratic in
$Fa/(k_B T)$ for $t \gg \tau_{1}$, since the linear terms cancel out.
Hence, we obtain the anomalous exponent in this limit
$\alpha_{\mathrm{w}}$ as follows:
\begin{align}
	\alpha_{\mathrm{w}}
	&\simeq
			\frac{2T}{T_c}	\ \ \  (t\gg \tau_1).
\label{12111203}			
\end{align}
This is coincident to the anomalous exponent for the purely diffusive
case \cite{ESW}.
Because the condition $\tau_1\sim\tau_2\sim\tau_3$ holds in this case,
both of the two nearest neighbors contribute to the anomalous exponent
at the same time scale, which is similar to the purely diffusive case.
As a result, the time evolution of the anomalous exponent is also
similar to that of the purely diffusive case.

\vspace{1em}
{\bf Strong field cases:}
Next, we consider the strong field limit where $\tau_1\ll \tau_3$ holds. 
In this limit, there are two situations according to the value of $T/T_c$:
(i)the strong disorder case ($T/T_c\ll 1$) where $\tau_1\leq  \tau_2\ll \tau_3$  holds
and 
(ii)the weak disorder case ($T/T_c\sim 1$) where $\tau_1\ll \tau_2\sim \tau_3$  holds. 
We investigate these two cases in the following.

First we consider the strong disorder case. 
In a time scale where $\tau_2\ll\ t \ll \tau_3$ holds, by substituting
Eqs.~(\ref{11071502}) and (\ref{approx}) into Eq.~(\ref{12110601}), we
obtain the effective waiting time in this limit $\langle w(t)
\rangle_{\mathrm{ss}}$ as follows:
\begin{align}
\langle w(t) \rangle_{\mathrm{ss}}
&\simeq
		\frac{1}{4}K_{1}
			e^{-\frac{Fa}{k_BT}}
			\frac{T}{T_c}(K_{1}e^{-\frac{Fa}{k_BT}}t)^{-1-\frac{T}{T_c}}			
			\Gamma\left(\frac{T}{T_c}+1\right)		
\nn
\\
&\propto t^{-1-\frac{T}{T_c}}.
\label{12111202}
\end{align}
Therefore, $\alpha_{\mathrm{ss}}\simeq T/T_c$. 
Next, in a time scale where $t\gg \tau_3$ holds, we obtain
\begin{align}
\langle w(t) \rangle_{\mathrm{ss}}
&\simeq
		\frac{1}{4}K_{1}
			e^{-\frac{Fa}{k_BT}}
			\left(\frac{T}{T_c}\right)^2(K_{1}e^{-\frac{Fa}{k_BT}}t)^{-1-\frac{T}{T_c}}
\nn
\\&\times			
			\Gamma\left(\frac{T}{T_c}+1\right)		
			(K_{1}t)^{-\frac{T}{T_c}}
			\Gamma\left(\frac{T}{T_c}\right)	
\nn
\\
&\propto t^{-1-2\frac{T}{T_c}},
\label{12111201}
\end{align}
which gives $\alpha_{\mathrm{ss}}\simeq 2T/T_c$. 
From Eqs.~(\ref{12111202}) and (\ref{12111201}), we obtain the
asymptotic expressions of the anomalous exponent in the strong field limit as
\begin{eqnarray}
	\alpha_{\mathrm{ss}}
	&\simeq&
	\left\{
		\begin{array}{cc}
			\frac{T}{T_c}	&	\left(\tau_3\gg  t \gg \tau_2\right)\\
			\frac{2T}{T_c}	& \left(t\gg \tau_3\right)
		\end{array}.
	\right.
	\label{12111204}
\end{eqnarray}
The time-dependent behavior in Eq.~(\ref{12111204}) is clearly different from that of the weak field case in Eq.~(\ref{12111203}).
This is caused by the strong external field.
If the strong external field is applied to the carriers, hops to the
opposite direction of the external field are suppressed and those in the
same direction are enhanced strongly. 
As a result, the anomalous behavior of carriers which hop in the direction of the external field appears first. 
This is the time region where $\alpha\simeq T/T_c$ holds. 
Next, the anomalous behavior of carriers which hop to the opposite
direction of the external field emerges. 
This is the time region where $\alpha\simeq 2T/T_c$ holds. 
{
It is worth mentioning that the microscopic spatial structure,
especially the discreteness of the hopping sites, is necessary for the
time scale separation and the discrete asymptotic values of the
anomalous exponent in Eq.~(\ref{12111204}). 
If the hopping sites distribute continuously, such a time-scale
separation does not occur and the anomalous exponent would not be
written in such simple approximate forms. 
Hence, the microscopic spatial structure of the hopping sites is
important even in one-dimensional systems.}

\vspace{1em}
It is also worth noting that $\tau_2\sim\tau_1$ holds for a finite $Fa/k_BT$ if $T/T_c$ is small enough.
It implies that the time scale where $\alpha_{ss}$ approaches $T/T_c$ is independent of  $Fa/k_BT$, if $T/T_c$ is small enough. 
On the contrary to this, since the time scale $\tau_3$ has no dependence on $T/T_c$, the time scale where $\alpha_{ss}$ approaches $2T/T_c$ is determined by  $Fa/k_BT$ solely.

\vspace{1em}
Secondly, we consider the weak disorder case where $\tau_1\ll \tau_2\sim \tau_3$ holds. 
This implies that the exponential terms in Eq.~(\ref{12110601}) become dominant. 
This is the usual Poisson distribution. 
Hence, the anomalous exponent $\alpha_{\mathrm{sw}}$ is given by
\begin{align}
	\alpha_{\mathrm{sw}}
	&\simeq
				1\ \left(t\gg \tau_3\right),
	\label{12111301}
\end{align}
i.e., the diffusion is normal.
These findings are confirmed in the next section.
%

{
As stated before, we focus on the systems with the exponential DOS
throughout this study. 
For other types of DOS, such as the Gaussian DOS, which is believed to
be typical to organic semiconductors, it is hard in general to obtain
simple asymptotic expressions of the anomalous exponent due to
mathematical difficulties.
However, the time-dependent behavior of the anomalous exponent can in
principle be obtained, even for these cases, by performing numerical
integrations of the effective waiting time.  }
 %

\section{Simulation}\label{Simulation}

\begin{table}
	\begin{tabular}{|c|c|c|c|c|c|}
		\hline
			Case 	&	\ $N_P$\ 	& \ $a/\xi$\ 	&	\ $\nu_0$ [sec$^{-1}$]\ 	& \ $T/T_c$\ 	& \ $Fa/k_BT$\ \\
		\hline
			\ref{Ano}-1	&	$10^6$	& $10$	&	$10^{-12}$	& $0.20$	& $0$\\
		\hline
			\ref{Ano}-2	&	$10^6$	& $10$	&	$10^{-12}$	& $0.20$	& $1$\\
		\hline
			\ref{SimW}	&	$10^6$	& $10$	&	$10^{-12}$	& $0.10$, $0.20$, $0.30$	& $1$\\		
		\hline
			\ref{SimS}-1	&	$10^5$	& $10$	&	$10^{-12}$	& $0.10$, $0.20$, $0.30$	& $15$\\		
		\hline
			\ref{SimS}-2	&	$10^5$	& $10$	&	$10^{-12}$	& $0.05$, $0.20$	& $15,\ 25$\\					
		\hline
			\ref{SimS}-3	&	$10^5$	& $10$	&	$10^{-12}$	& $0.20$, $0.40$, $0.60$, $0.80$	& $10$\\		
		\hline
	\end{tabular}
	\caption{Summary of simulation conditions: 
	The parameter $N_P$ is the number of carriers which  is
 essentially the number of the trials of the simulation performed. 
	Other parameters are defined in the text. 
	All the carriers are initially rested at the origin for all cases. }
	\label{SimCon}
\end{table}	

\subsection{Anomalous behaviors}\label{Ano}
The approximate forms of the anomalous exponent in the weak and strong
external field limits given by Eqs.~(\ref{12111203}), (\ref{12111204})
and (\ref{12111301}), are examined by the MC simulation of the hopping
conductance.  
Throughout this article, we study a one-dimensional system with carriers
allowed to hop up to the nearest neighboring sites, which is shown in
Fig.~\ref{system}.

Before examining Eqs.~(\ref{12111203}), (\ref{12111204}), and
(\ref{12111301}), we confirm that the time evolution of the spatial
distribution of the carriers shows anomalous behaviors such as anomalous
diffusion and anomalous advection-diffusion.
{
We adopt the numerical algorithm of Ref.~\cite{ESW}, where hoppings of
the walkers are synchronized.
%
%

In Fig.~\ref{behavior}, we show the time evolution of the spatial
distribution of the carriers.  
The horizontal axes are the dimensionless positions of the carriers,
which are normalized by the lattice spacing, and the vertical axes are
the numbers of the carriers.  
The three symbols of figures (triangle, rectangle, and circle)
correspond to three different times.
From the upper figure of Fig.~\ref{behavior}, one can see that the
carriers exhibit anomalous diffusion if no external field is applied.
From the lower figure of Fig.~\ref{behavior}, one can see that the
carriers exhibit anomalous diffusion-advection if an external field is
applied.
Thus we have verified that the carriers show the anticipated anomalous
behaviors.
The conditions of the simulations are collected in ``Case \ref{Ano}-1''
and ``Case \ref{Ano}-2'' of TABLE~\ref{SimCon} for the anomalous
diffusion 
and the anomalous advection-diffusion,
%
respectively.

The process we are considering consists of many small random steps.
Each small random step is a hopping to the nearest neighbors. 
As a result of many random steps, the carriers diffuse and advect, and
the statistical long-time behavior of the carriers is obtained.

%
\begin{figure}[htb]
\includegraphics[width=7cm]{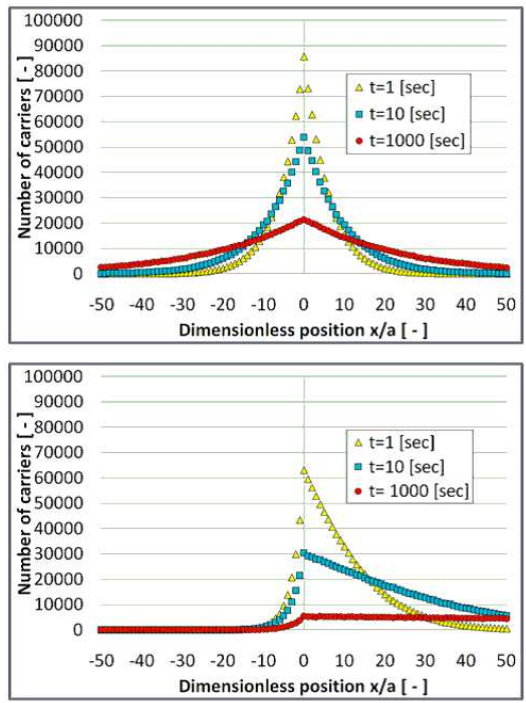}	
\caption{The time evolution of the spatial distribution of the carriers,
 for the case with (lower figure) and without (upper figure) an external
 field. 
The carriers exhibit anomalous diffusion when no external field is
 applied, while they exhibit anomalous diffusion-advection when an external field is applied.
 The conditions of the simulation are shown in ``Case \ref{Ano}-1''  
 (upper figure) and ``Case \ref{Ano}-2'' (lower figure) in TABLE \ref{SimCon}, respectively.}
\label{behavior}
\end{figure}
%


\subsection{Weak field cases}\label{SimW}

Now, we examine the expression of the anomalous exponent in Eq.~(\ref{12111203}). 
In order to estimate the anomalous exponent, we have used the following relation \cite{PhysRep.339.1}
\begin{eqnarray}
 \langle x(t) \rangle_{F}\propto t^{\alpha_F}.
\label{xF}
\end{eqnarray}
Here, $\langle x(t) \rangle_{F}$ is the mean displacement and $\alpha_F$ is the anomalous exponent for the case when the external field is applied to the carriers. 
Eq.~(\ref{xF}) holds if all the carriers are rested at the origin initially. 
From Eq.~(\ref{xF}), we can estimate $\alpha_F$ by fitting the simulation results at time $t$ to the function $f(t)\propto t^{\alpha}$.
{
In addition, we also consider the mean velocity of carriers $\langle
v(t)\rangle$, to which the carrier current $I(t)$ is proportional, i.e.,
$I(t)\propto \langle v(t)\rangle\equiv d\langle x(t)\rangle/dt$.
It is also known that $\langle v(t)\rangle\propto t^{-1+\alpha}$ and
$\langle v(t)\rangle\propto t^{-1-\alpha}$ holds for short and long time
scales, respectively, when the absorbing boundary condition is
implemented to the system.
This behavior explains well the TOF signal 
\cite{Xerox1,PhysRevE.63.046118}.
In the present cases, the absorbing boundary condition is not
implemented. Hence, $\langle v(t)\rangle$ is proportional to
$t^{-1+\alpha}$.  }
%

%
\begin{figure}[htb]
	\subfigure[Time evolution of the mean velocity]
	{
		\includegraphics[width=7cm]{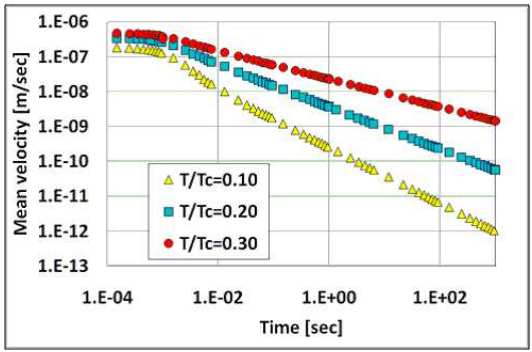}	
		\label{weak1_v}
	}
	\subfigure[Time evolution of the anomalous exponent]
	{
		\includegraphics[width=7cm]{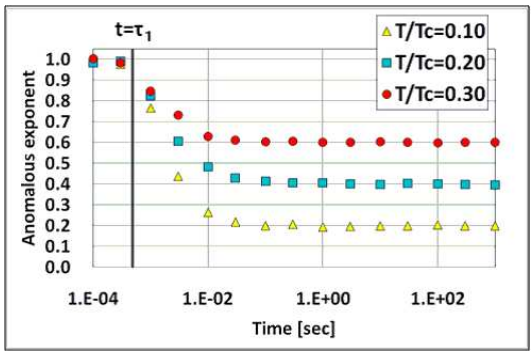}	
		\label{weak1}
	}
\caption{ The time evolution of the mean velocity (Fig.~\ref{weak1_v})
 and the anomalous exponent (Fig.~\ref{weak1}) for the case of a weak field. 
The anomalous exponent for the case of a weak field coincides with that
 for the diffusive case.
The conditions of the simulation are shown in ``Case \ref{SimW}'' in TABLE \ref{SimCon}.  
The value of $\tau_1$ is equal to $4.85\times 10^{-4}\ \mathrm{[sec]}$.
}
\end{figure}

We show the result of the MC simulation for the time evolution of the
mean velocity (Fig.~\ref{weak1_v}) and the anomalous exponent (Fig.~\ref{weak1}).
The condition of the simulation is shown in ``Case \ref{SimW}'' in
TABLE \ref{SimCon}.  
We have chosen $Fa/k_BT=1$ so that the condition $\tau_1\sim\tau_2\sim\tau_3$ holds. 
Then, the value of $\tau_1$ is equal to $4.85\times 10^{-4}\ \mathrm{[sec]}$. 
%
The horizontal axes are the time and the vertical axes are the mean
velocity (Fig.~\ref{weak1_v}) and the anomalous exponent
(Fig.~\ref{weak1}), respectively.
The three symbols in the figures (triangle, rectangle, and circle)
correspond to three different values of $T/T_c$.
%
%
From Fig.~\ref{weak1}, one can see that the anomalous exponent is
compatible with Eq.~(\ref{12111203}).  
%

\subsection{Strong field cases}\label{SimS}	       

We consider the strong field limit. 
First, we examine the validity of Eq.~(\ref{12111204}). 
Then, we examine that of Eq.~(\ref {12111301}). 

We show the result of the MC simulation for the time evolution  of the
mean velocity (Fig.~\ref{transition_v}) and the anomalous exponent
(Fig.~\ref{transition}). 
The condition of the simulation is shown in ``Case \ref{SimS}-1'' in
TABLE \ref{SimCon}.
We have chosen  $Fa/k_BT=15$ so that the condition $\tau_1\leq \tau_2\ll \tau_3$ holds. 
Then, the values of $\tau_1$ and $\tau_3$ are equal to $4.85\times 10^{-4}\ \mathrm{[sec]}$ and 
$1.59\times 10^{3}\ \mathrm{[sec]}$, respectively.
%
The horizontal axes are the time and the vertical axes are the mean
velocity and the anomalous exponent, respectively.
The three symbols in the figures (triangle, rectangle, and circle)
correspond to three different values of $T/T_c$.
	      
From Fig.~\ref{transition}, one can see that the values of anomalous
exponent first decrease from 1 to $T/T_c$, and then increase up to
$2T/T_c$.
Moreover, one can also see that the time scales where $\alpha$ reaches
$2T/T_c$ are independent of the value of $T/T_c$.  
\begin{figure}[htb]
	\subfigure[Time evolution of the mean velocity]
	{
		\includegraphics[width=7cm]{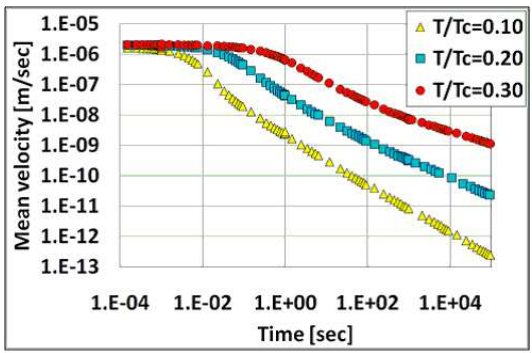}	
		\label{transition_v}
	}
	\subfigure[Time evolution of the anomalous exponent]
	{
		\includegraphics[width=7cm]{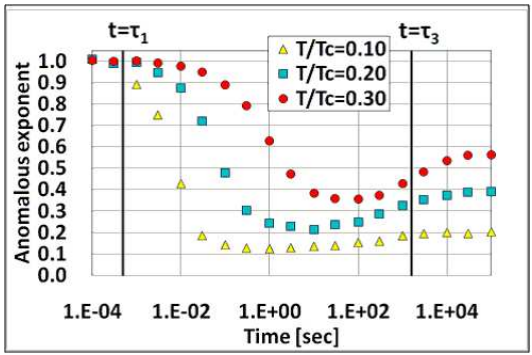}	
		\label{transition}
	}
\caption{ The time evolution of the mean velocity (Fig.~\ref{transition_v}) and the anomalous exponent (Fig.~\ref{transition}) for the case of a strong field. 
The conditions of the simulation are in ``Case \ref{SimS}-1'' in TABLE \ref{SimCon}.  
The values of $\tau_1$ and $\tau_3$ are equal to $4.85\times 10^{-4}\ \mathrm{[sec]}$ and 
$1.59\times 10^{3}\ \mathrm{[sec]}$, respectively.}

\end{figure}

\vspace{1em}			
%
\begin{figure}[htb]
	\subfigure[Time evolution of the mean velocity]
	{
		\includegraphics[width=7cm]{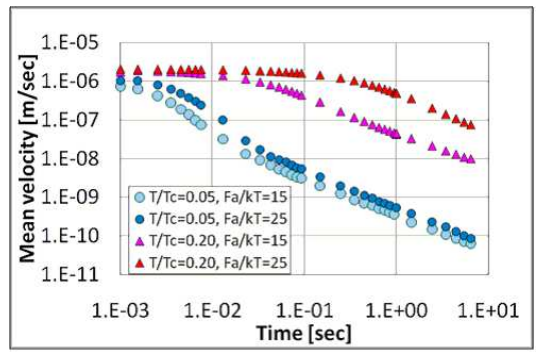}	
		\label{tau2_v}
	}
	\subfigure[Time evolution of the anomalous exponent]
	{
		\includegraphics[width=7cm]{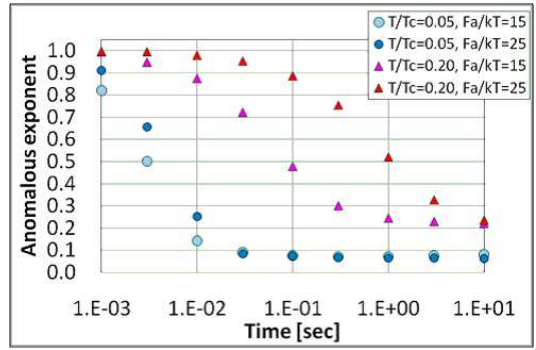}	
		\label{tau2}
	}

\caption{The time evolution of the mean velocity (Fig.~\ref{tau2_v})  and the anomalous exponent
 (Fig.~\ref{tau2}) for the case of a strong field and a strong disorder. 
 The time required for the anomalous exponent $\alpha$ to approach
 $T/T_c$ is almost independent of the value of $Fa/k_BT$ for the case of
 $T/T_c=0.05$, while it is not for the case of $T/T_c=0.20$. 
The conditions of the simulation are shown in ``Case \ref{SimS}-2'' in TABLE \ref{SimCon}.  
}

\end{figure}
In the previous section, it is also predicted that the time scale where
$\alpha_{\mathrm{ss}}$ approaches $T/T_c$ is independent of $Fa/k_BT$,
if $T/T_c$ is small enough.
In order to confirm this, we performed simulations, where the conditions
are shown in ``Case \ref{SimS}-2'' in TABLE \ref{SimCon}.
We have chosen  $T/T_c=0.05$, $0.2$ and $Fa/k_BT=15$, $25$. 
For the case $T/T_c=0.05$, $\tau_2 = 1.03\times 10^{-3}$ [sec$^{-1}$]
for $Fa/k_BT=15$ and $\tau_2 = 1.69\times 10^{-3}$ [sec$^{-1}$] for $Fa/k_BT=25$. 
For the case $T/T_c=0.2$, $\tau_2 = 2.06\times 10^{-2}$ for $Fa/k_BT=15$
and $\tau_2 = 2.51\times 10^{-1}$ for $Fa/k_BT=25$. 
	    
We show the results of the MC simulation for the time evolutions of the
mean velocity (Fig.~\ref{tau2_v}) and the anomalous exponent
(Fig.~\ref{tau2}).  
%
%
The four symbols in the figures correspond to four different pairs of $T/T_c$
and $Fa/k_BT$.  One can see that the time scale where $\alpha_{\mathrm{ss}}$
approaches $T/T_c$ is almost independent of the value of $Fa/k_BT$ for
the cases of $T/T_c=0.05$. On the contrary, it depends on the value of
$Fa/k_BT$ for the cases of $T/T_c=0.20$, as expected. This is consistent
with the theoretical prediction.  
			
\vspace{1em}
%
\begin{figure}[htb]

	\subfigure[Time evolution of the mean velocity]
	{
		\includegraphics[width=7cm]{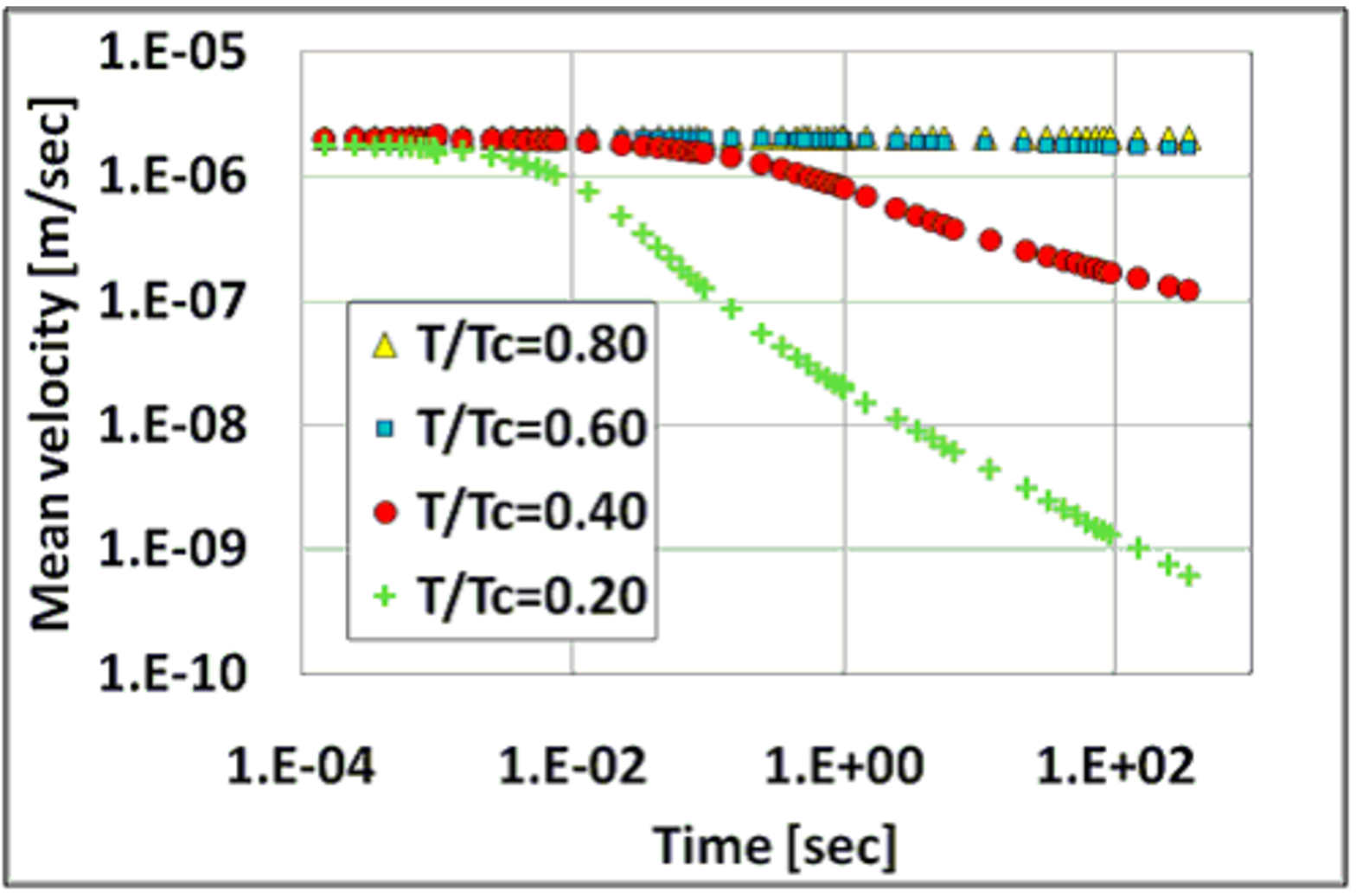}	
		\label{diffusion_v}
	}
	\subfigure[Time evolution of the anomalous exponent]
	{
		\includegraphics[width=7cm]{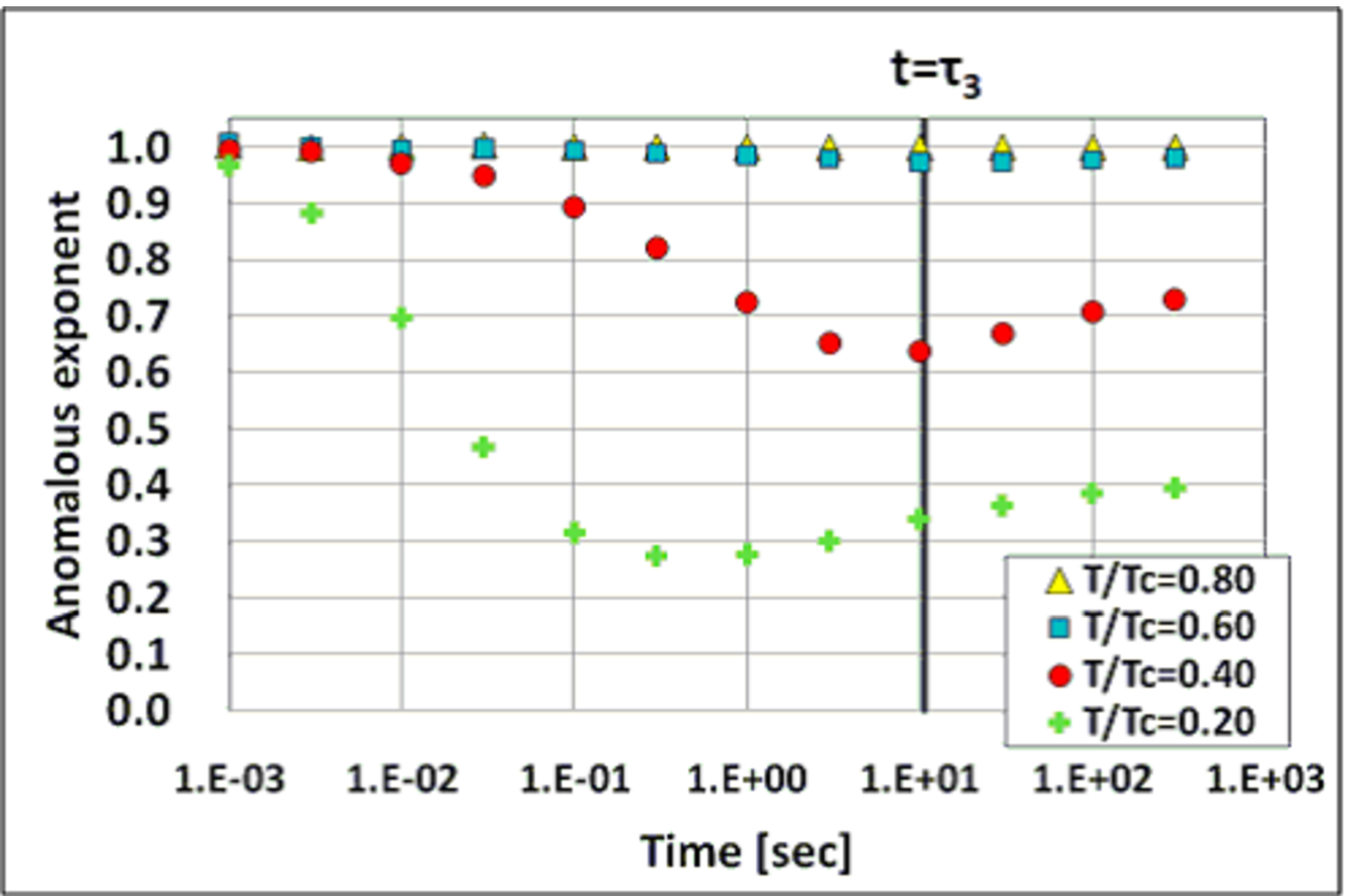}	
		\label{diffusion}
	}

\caption{ The time evolutions of the mean velocity
 (Fig.~\ref{diffusion_v}) and the anomalous exponent
 (Fig.~\ref{diffusion}) for the case of a strong field and a weak disorder. 
 The anomalous exponent approaches $1$ as $T/T_c$ increases. 	      
 The conditions of the simulation are shown in ``Case \ref{SimS}-3'' in TABLE \ref{SimCon}.  
}

\end{figure}
Finally, we confirm the validity of Eq.~(\ref{12111301}). 
%
We show the results of the MC simulation for the time evolutions of the
mean velocity (Fig.~\ref{diffusion_v}) and the anomalous exponent
(Fig.~\ref{diffusion}).  The conditions of the simulation are shown in ``Case
\ref{SimS}-3'' in TABLE \ref{SimCon}.  
We have set $Fa/k_BT = 10$ so that the condition $\tau_1\ll
\tau_3$ holds.  Then, $\tau_1 = 4.85\times 10^{-4}\ \mathrm{[sec]}$ and $\tau_3 = 1.07\times 10^{1}\
\mathrm{[sec]}$, respectively.
		
The horizontal axes are the time and the vertical axes are the mean velocity and the anomalous exponent,
respectively. 
The three symbols in the figures (triangle, rectangle, and circle) correspond
to three different values of $T/T_c$.  We also show the results of
$T/T_c=0.2$ (cross) for reference.  One can see that the anomalous
exponent approaches $1$ as $T/T_c$ increases. 
%

\section{Discussion}\label{Discussion}
In this section, we discuss the relation of our results to previous studies for both the weak and strong field cases.

For the weak field cases, Barkai et
al. \cite{PhysRevE.61.132,PhysRevE.63.046118} have shown that CTRW with
the anomalous exponent $\alpha$ is described by the fractional
Fokker-Planck equation (FFPE) and the FFPE describes the behavior of the
SMM qualitatively.  Because the disorder model can be regarded as a CTRW
model with the anomalous exponent of Eq.~(\ref{12111203}), by the coarse
graining performed above, it is expected that the disorder model in this
limit is described by the FFPE and its behavior is qualitatively
consistent with the SMM.  Moreover, although $\alpha$ is a parameter in
their theory, we have derived it from a microscopic physical
model. Therefore, our results make it possible to estimate the transport
coefficients in the FFPE quantitatively, and to compare the results of
the simulations with experiments or other models, such as the SMM,
quantitatively. These are future works.

\vspace{1em}
It is also worth noting that Richert et al. \cite{PhysRevLett.63.547} have shown that the Einstein relation  \cite{AnnPhys.17.549}
\begin{align}	
	\frac{D}{\mu}=\frac{k_BT}{e}
	\label{13041001}
\end{align}
 does not hold in the disorder model. 
 Here, $D$ is the diffusion constant and $\mu$ is the mobility.
They have shown that the disorder model shows a deviation from the
Einstein relation and the magnitude of the deviation increases as time
evolves.

On the other hand, Barkai et
al. \cite{PhysRevE.58.1296,PhysRevE.63.046118} have shown the
following features. 
If the anomalous exponent of the waiting time of a system with and
without a driving force, $\alpha_F$ and $\alpha_0$, are the same, i.e.,
the condition
\begin{align}
\alpha_0&=\alpha_F=\alpha
\label{13040201}
\end{align}
holds, the corresponding mean square displacements, $\langle
x(t)\rangle_F$ and $\langle x^2(t)\rangle_0$, are proportional to
$t^\alpha$,
and the generalized Einstein relation 
\begin{align}
\langle x^2(t)\rangle_0=2\frac{k_BT}{F}\langle x(t)\rangle_F
\label{13041002}
\end{align}
holds instead of Eq.~(\ref{13041001}) \cite{PhysRevE.58.1296}. 
%
Moreover,  Barkai \cite{PhysRevE.63.046118} has conjectured that the
generalized Einstein relation would hold for the hopping conduction by
assuming that the condition $\alpha_0=\alpha_F$ holds. 
Because we have shown that Eq.~(\ref{13040201}) holds 
for the disorder model in the weak field limit, our results support
their assumptions. 
Therefore, the generalized Einstein relation Eq.~(\ref{13041002}),
rather than Eq.~(\ref{13041001}), holds in the disorder model.

\vspace{1em}
For the strong field and strong disorder cases, it is important that
$\alpha$ takes the value of $T/T_c$, which is the same as that of the
MTM, before approaching $2T/T_c$. 
This implies that there is a time region in this case that the anomalous
exponent of the disorder model coincides with that of the MTM. 
The relation of the disorder model to the MTM has been studied by
Hartenstein et al. \cite{HBJK}.  
They have {\it numerically} shown that the waiting times of both models
are similar if $\alpha$ of both models are equal. In this study, we have
{\it theoretically} found a time region where $\alpha$ of both models
are equal. 
%

{
Throughout this study, we have mainly focused on the one-dimensional
system with a DC electric field. 
It is worth noting that the extrapolation of our analytical results to
three-dimensional systems would not be straightforward, because the
behaviors of the hopping conduction depend on the dimensionality of the
system in the presence of an external field. 
%
%
In addition, it is significant to analyze systems with AC fields and
compare the result with previous studies, such as the coherent medium
approximation and the bond percolation model
\cite{PhysRevB.24.5284,PhysRevB.26.6480}.
These are future works.  }

		\section{Summary}\label{Summary}
In this study, we have theoretically investigated the DC electric field
effect on the anomalous exponent of the hopping conduction in the
disorder model.
First of all, we have calculated the effective waiting time of a general
system with an external field and shown that the time evolution of the
anomalous exponent depends on the external field.  
We have found that there are three typical time scales for each pair of
sites, which are (1) the typical time scale of the diffusion, (2) the
time scale where the anomalous behavior of the hopping in the direction
of the field appears, which depends on the strength of the disorder and
the external field, and (3) the time scale where the anomalous behavior
of the hopping in the {\it opposite} direction of the field appears,
which is independent of the strength of the disorder.

Next, we have focused on a one-dimensional system where carriers can hop
only to the nearest neighbors, in order to illustrate the features of
the anomalous exponent, and calculated the effective waiting time of
this system.
For the case of a weak field, we have shown that the anomalous
exponent is given by $2T/T_c$ and is consistent with that of diffusive
systems. 
For the case of a strong field, we have considered strong and weak
disorder cases.
For the strong field and the strong disorder case, the time evolution of
the anomalous exponent clearly differs from that for the case of a weak
field.
The anomalous exponent first decreases from 1 to $T/T_c$, and then
increases up to $2T/T_c$.
We have also shown that the time scale where $\alpha_{\mathrm{ss}}$
approaches $T/T_c$ is independent of $Fa/k_BT$, if $T/T_c$ is small
enough.
In contrast, the time scale where $\alpha_{\mathrm{ss}}$ approaches
$2T/T_c$ is determined by $Fa/k_BT$ solely.
For the strong field and the weak disorder case, the exponent is equal
to 1. 
It is reasonable that the transport is normal in the weak disorder
limit.

We have verified the above theoretical predictions by means of MC
simulation of the hopping conductance.  
We have chosen a one-dimensional system with carriers allowed to hop
only to the nearest neighboring sites.  
The hopping process consists of many small random steps, which results
in the statistical long-time behavior of the carriers, i.e., anomalous
diffusion and advection.
We have demonstrated that the carriers actually show the expected
anomalous behavior.

Then, we have verified the effect of an external field on the anomalous
exponent.  
For the case of a weak field, we have verified that the anomalous
exponent is consistent with that of diffusive systems.
For the case of a strong field and a strong disorder, we have verified
that the anomalous exponent first decreases from 1 to $T/T_c$, and then
increases up to $2T/T_c$ in the long time limit.
Characteristic time scales of these behaviors are consistent with the
theoretical predictions. 
For the case of a strong field and a weak disorder, we have verified
that the value of the anomalous exponent is equal to 1.  
%



\vspace{1em}
{\it Acknowledgements}.
We are grateful to Mr. Shinjo, Mr. Nagane, and the members of the Analysis Technology Development Department 1 for 
fruitful discussions and 
their support.
%
	\appendix
	\begin{widetext}
	\section{Derivation of Eq.~(\ref{11060902})}\label{Appendix}
In this appendix, we derive Eq.~(\ref{11060902}). In the following we consider the case $T/T_c<1$ and $F\geq 0$ which we are interested in this study. For other cases, e.g. $F\leq 0$, similar techniques would be applicable.
We start from the Eq.~(\ref{wtev})
\begin{align}
\langle w(t) \rangle_F
&=
	\left(\prod_{j\in \mathcal{N}}\int_{-\infty}^\infty\!\! d \epsilon_{ji}
	\frac{e^{-\frac{|\ep_{ji}|}{k_BT_c}}}{2k_BT_c}
	\right)
	\left(
			\sum_{k\in\mathcal{N}}\!\! K_{ki}
			e^{
				-\frac{(\epsilon_{ki}-Fx_{ki})\Theta(\epsilon_{ki}-Fx_{ki})}{k_BT}}
			e^{
				-\sum_{l\in\mathcal{N}} K_{li}
			e^{-\frac{(\epsilon_{li}-Fx_{li})\Theta(\epsilon_{li}-Fx_{li})}{k_BT}} t
			}
	\right).
	\label{wtev2}
\end{align}
We can calculate the integrals in Eq.~(\ref{wtev2}) by two successive transformations of the integration variables \cite{US}.

We divide $\mathcal{N}$ into three sets $\mathcal{N}_+$, $\mathcal{N}_0$ and $\mathcal{N}_-$, according to the sign of $x_{li}$.
Here, $\mathcal{N}_+$ represents a set of neighboring sites which satisfy the condition $x_{li}>0$, $\mathcal{N}_0$ represents that of neighboring sites which satisfy the condition $x_{li}=0$, and 
$\mathcal{N}_-$ represents that of neighboring sites which satisfy $x_{li}<0$.

First, we consider the case $x_{li}>0$. In this case, we can calculate the integral as follows:
\begin{align}
		\int_{-\infty}^\infty d \epsilon_{li}
		\frac{e^{-\frac{|\ep_{li}|}{k_BT_c}}}{2k_BT_c}
		e^{
			-K_{li}
		e^{-\frac{(\epsilon_{li}-Fx_{li})\Theta(\epsilon_{li}-Fx_{li})}{k_BT}} t
		}
	=
		\int_{-\infty}^{Fx_{li}} d \epsilon_{li}
		\frac{e^{-\frac{|\ep_{li}|}{k_BT_c}}}{2k_BT_c}
		e^{
			-K_{li}t
		}
		+
		\int_{Fx_{li}}^\infty d \epsilon_{li}
		\frac{e^{-\frac{\ep_{li}}{k_BT_c}}}{2k_BT_c}
		e^{
			- K_{li}
		e^{-\frac{\epsilon_{li}-Fx_{li}}{k_BT}} t
		}.
		\label{12102901}
\end{align}
The first term in Eq.~(\ref{12102901}) can be calculated as
\begin{align}
		\int_{-\infty}^{Fx_{li}} d \epsilon_{li}
		\frac{e^{-|\ep_{li}|/k_BT_c}}{2k_BT_c}
		e^{
			- K_{li}t
		}
	&=
		e^{
			- K_{li}t
		}		
		-
		\frac{1}{2}e^{-\frac{Fx_{li}}{k_BT}\frac{T}{T_c}}
		e^{
			- K_{li}t
		}.		
		\label{12102902}
\end{align}
The second term in Eq.~(\ref{12102901}) can be calculated by two successive transformations of the integration variables, i.e.,  $A_{li}=e^{-(\epsilon_{li}-Fx_{li}) /k_B T}$ for the first step, and $C_{li}=K_{li} A_{li} t$ for the second. 
By the first transformation $A_{li}=e^{-(\epsilon_{li}-Fx_{li}) /k_B T}$, and using $d A_{li}/d \epsilon_{li} =-A_{li}/{k_BT}$, we obtain
\begin{align}
		\int_{Fx_{li}}^\infty d \epsilon_{li}
		\frac{e^{-\frac{\ep_{li}}{k_BT_c}}}{2k_BT_c}
		e^{
			-K_{li}
		e^{-(\epsilon_{li}-Fx_{li})/k_BT} t
		}
	&=
		\frac{1}{2}e^{-\frac{Fx_{li}}{k_BT}\frac{T}{T_c}}
		\frac{T}{T_c}\int_{0}^1 dA_{li} 
		A_{li}^{-1+\frac{T}{T_c}}
		e^{-K_{li}A_{li}t}.
\end{align}
Then, by the second transformation $C_{li}=K_{li} A_{li} t$, we obtain
\begin{align}
	\frac{1}{2}e^{-\frac{Fx_{li}}{k_BT}\frac{T}{T_c}}
	\frac{T}{T_c}\int_{0}^1 dA_{ji} 
		A_{ji}^{-1+\frac{T}{T_c}}
		e^{-K_{ji}A_{ji}t}
	&=
	\frac{1}{2}e^{-\frac{Fx_{li}}{k_BT}\frac{T}{T_c}}
		\frac{T}{T_c}(K_{li}t)^{-\frac{T}{T_c}}
		\gamma\left(\frac{T}{T_c},K_{li}t\right).		
		\label{11072901}
\end{align}		
Here, 
\begin{align}
	\gamma(T/T_c,K_{ij}t)\equiv \int_0^{K_{ij}t}d \tau
		\tau^{-1+T/T_c}e^{-\tau}	
\end{align}
 is the lower incomplete gamma function.
 Substituting Eqs.~(\ref{11072901}) and (\ref{12102902}) into Eq.~(\ref{12102901}), we obtain
\begin{align}
		\int_{-\infty}^\infty d \epsilon_{li}
		\frac{e^{-|\ep_{li}|/k_BT_c}}{2k_BT_c}
		e^{
			-K_{li}
		e^{-(\epsilon_{li}-Fx_{li})\Theta(\epsilon_{li}-Fx_{li})/k_BT} t
		} 
	&=
		\left(
		1
		-
		\frac{1}{2}e^{-\frac{Fx_{li}}{k_BT}\frac{T}{T_c}}
		\right)
		e^{
			- K_{li}t
		}	
	+
	\frac{1}{2}e^{-\frac{Fx_{li}}{k_BT}\frac{T}{T_c}}
		\frac{T}{T_c}(K_{li}t)^{-\frac{T}{T_c}}
		\gamma\left(\frac{T}{T_c},K_{li}t\right)	
	\nn\\&\equiv I^{(0)+}_{li}.						
		\label{12102910}
\end{align}
Similarly, we obtain
\begin{align}
	&
		\int_{-\infty}^\infty d \epsilon_{li}
		\frac{e^{-|\ep_{li}|/k_BT_c}}{2k_BT_c}
		e^{-\frac{(\epsilon_{li}-Fx_{li})\Theta(\epsilon_{li}-Fx_{li})}{k_BT}}
		e^{
			-K_{li}
		e^{-(\epsilon_{li}-Fx_{li})\Theta(\epsilon_{li}-Fx_{li})/k_BT} t
		} 
	\nn\\
	&=
		\left(
		1
		-
		\frac{1}{2}e^{-\frac{Fx_{li}}{k_BT}\frac{T}{T_c}}
		\right)
		e^{
			- K_{li}t
		}	
	+
	\frac{1}{2}e^{-\frac{Fx_{li}}{k_BT}\frac{T}{T_c}}
		\frac{T}{T_c}(K_{li}t)^{-1-\frac{T}{T_c}}
		\gamma\left(\frac{T}{T_c}+1,K_{li}t\right)
	\equiv I^{(1)+}_{li}.			
		\label{11072902}
\end{align}

Next, we consider the case $x_{li}= 0$. For this case, we can adapt the above mathematical techniques again:
\begin{align}
		\int_{-\infty}^\infty d \epsilon_{li}
		\frac{e^{-|\ep_{li}|/k_BT_c}}{2k_BT_c}
		e^{
			-K_{li}
		e^{-(\epsilon_{li})\Theta(\epsilon_{li})/k_BT} t
		}
	&
	=
		\int_{-\infty}^{0} d \epsilon_{li}
		\frac{e^{-|\ep_{li}|/k_BT_c}}{2k_BT_c}
		e^{
			-K_{li}t
		}
		+
		\int_{0}^\infty d \epsilon_{li}
		\frac{e^{-|\ep_{li}|/k_BT_c}}{2k_BT_c}
		e^{
			- K_{li}
		e^{-\epsilon_{li}/k_BT} t
		}.
		\label{12112601}
\end{align}
The first term in Eq.~(\ref{12112601}) can be calculated as
\begin{align}
		\int_{-\infty}^{0} d \epsilon_{li}
		\frac{e^{-|\ep_{li}|/k_BT_c}}{2k_BT_c}
		e^{
			-K_{li}t
		}
	&=
	\frac{1}{2}
	e^{
			-K_{li}t
		}.
		\label{12112602}
 \end{align} 
Using the variable transformations in the above, the second term in Eq.~(\ref{12112601}) can be calculated as follows:
 \begin{align}
		\int_{0}^\infty d \epsilon_{li}
		\frac{e^{-\ep_{li}/k_BT_c}}{2k_BT_c}
		e^{
			- K_{li}
		e^{-\epsilon_{li}/k_BT} t
		}			
	&=
		\frac{1}{2}\frac{T}{T_c}
		(K_{li}t)^{-\frac{T}{T_c}}
		\gamma\left(\frac{T}{T_c},K_{li}t\right).
		\label{12112603}
\end{align}
 
Substituting Eqs.~(\ref{12112602}) and (\ref{12112603}) into Eq.~(\ref{12112601}), we obtain
 \begin{align}
		\int_{-\infty}^\infty d \epsilon_{li}
		\frac{e^{-|\ep_{li}|/k_BT_c}}{2k_BT_c}
		e^{
			-K_{li}
		e^{-\epsilon_{li}\Theta(\epsilon_{li})/k_BT} t
		}
	&=
		\frac{1}{2}
	e^{
			-K_{li}t
		}			
	+
		\frac{1}{2}\frac{T}{T_c}
		(K_{li}t)^{-\frac{T}{T_c}}
		\gamma\left(\frac{T}{T_c},K_{li}t\right)
	\equiv I^{(0)0}_{li}.					
		\label{12112605}
\end{align}
 
Similarly, we obtain 
 \begin{align}
		\int_{-\infty}^\infty d \epsilon_{li}
		\frac{e^{-|\ep_{li}|/k_BT_c}}{2k_BT_c}
		e^{-\epsilon_{li}\Theta(\epsilon_{li})/k_BT} 
		e^{
			-K_{li}
		e^{-\epsilon_{li}\Theta(\epsilon_{li})/k_BT} t
		}
	&=
		\frac{1}{2}
	e^{
			-K_{li}t
		}	
	+
		\frac{1}{2}
		(K_{li}t)^{-1-\frac{T}{T_c}}
		\gamma\left(\frac{T}{T_c}+1,K_{li}t\right)
	\equiv I^{(1)0}_{li}.					
		\label{12112606}
\end{align}

Finally, we consider the case $x_{li}< 0$. For this case, we can adapt the above mathematical techniques again:
\begin{align}
	&
		\int_{-\infty}^\infty d \epsilon_{li}
		\frac{e^{-|\ep_{li}|/k_BT_c}}{2k_BT_c}
		e^{
			-K_{li}
		e^{-(\epsilon_{li}-Fx_{li})\Theta(\epsilon_{li}-Fx_{li})/k_BT} t
		}
	\nn\\
	&=
		\int_{-\infty}^{Fx_{li}} d \epsilon_{li}
		\frac{e^{-|\ep_{li}|/k_BT_c}}{2k_BT_c}
		e^{
			-K_{li}t
		}
		+
		\int_{Fx_{li}}^\infty d \epsilon_{li}
		\frac{e^{-|\ep_{li}|/k_BT_c}}{2k_BT_c}
		e^{
			- K_{li}
		e^{-(\epsilon_{li}-Fx_{li})/k_BT} t
		}.
		\label{12102903}
\end{align}
The first term in Eq.~(\ref{12102903}) can be calculated as
\begin{align}
		\int_{-\infty}^{Fx_{li}} d \epsilon_{li}
		\frac{e^{-|\ep_{li}|/k_BT_c}}{2k_BT_c}
		e^{
			-K_{li}t
		}
	&=
	\frac{1}{2}e^{Fx_{li}/k_BT_c}
	e^{
			-K_{li}t
		}.
		\label{12102907}
 \end{align} 
The second term in Eq.~(\ref{12102903}) can be calculated as 
\begin{align}
	&
		\int_{Fx_{li}}^\infty d \epsilon_{li}
		\frac{e^{-|\ep_{li}|/k_BT_c}}{2k_BT_c}
		e^{
			- K_{li}
		e^{-(\epsilon_{li}-Fx_{li})/k_BT} t
		}
	\nn\\
	&=
		\int_{Fx_{li}}^0 d \epsilon_{li}
		\frac{e^{\ep_{li}/k_BT_c}}{2k_BT_c}
		e^{
			- K_{li}
		e^{-(\epsilon_{li}-Fx_{li})/k_BT} t
		}
		+
		\int_{0}^\infty d \epsilon_{li}
		\frac{e^{-\ep_{li}/k_BT_c}}{2k_BT_c}
		e^{
			- K_{li}
		e^{-(\epsilon_{li}-Fx_{li})/k_BT} t
		}	.		
		\label{12102904}
\end{align} 
Using the variable transformations in the above, the first term in Eq.~(\ref{12102904}) can be calculated as
\begin{align}
	&
		\int_{Fx_{li}}^0 d \epsilon_{li}
		\frac{e^{\ep_{li}/k_BT_c}}{2k_BT_c}
		e^{
			- K_{li}
		e^{-(\epsilon_{li}-Fx_{li})/k_BT} t
		}
	\nn\\
	&=
			-\frac{1}{2}e^{\frac{Fx_{li}}{k_BT}\frac{T}{T_c}}		
			e^{-K_{li}t}			
			+			
			\frac{1}{2}e^{\frac{Fx_{li}}{k_BT}\frac{T}{T_c}\left(1-\frac{T}{T_c}\right)}
			e^{-K_{li}e^{\frac{Fx_{li}}{k_BT}\frac{T}{T_c}}t}				
			+
			\frac{1}{2}e^{\frac{Fx_{li}}{k_BT}\frac{T}{T_c}}
		(K_{li}t)^{\frac{T}{T_c}}\gamma\left(1-\frac{T}{T_c},K_{li}t\right)
		\nn\\
		&
			-
			\frac{1}{2}e^{\frac{Fx_{li}}{k_BT}\frac{T}{T_c}}
		(K_{li}t)^{\frac{T}{T_c}}\gamma\left(1-\frac{T}{T_c},K_{li}e^{\frac{Fx_{li}}{k_BT}\frac{T}{T_c}}t\right).
		\label{12102905}
\end{align}

The second term in Eq.~(\ref{12102904}) can be calculated as
 \begin{align}
		\int_{0}^\infty d \epsilon_{li}
		\frac{e^{-\ep_{li}/k_BT_c}}{2k_BT_c}
		e^{
			- K_{li}
		e^{-(\epsilon_{li}-Fx_{li})/k_BT} t
		}			
	&=
		\frac{1}{2}e^{-\frac{Fx_{li}}{k_BT}\frac{T}{T_c}}
		\frac{T}{T_c}(K_{li}t)^{-\frac{T}{T_c}}
		\gamma\left(\frac{T}{T_c},K_{li}e^{\frac{Fx_{li}}{k_BT}}t\right).
		\label{12112101}
\end{align}
 
Substituting Eqs.~(\ref{12102907}), (\ref{12102905}), and (\ref{12112101}
)   into Eq.~(\ref{12102903}), we obtain
 \begin{align}
	&
		\int_{-\infty}^\infty d \epsilon_{li}
		\frac{e^{-|\ep_{li}|/k_BT_c}}{2k_BT_c}
		e^{
			-K_{li}
		e^{-(\epsilon_{li}-Fx_{li})\Theta(\epsilon_{li}-Fx_{li})/k_BT} t
		}
	\nn\\
	&=
			\frac{1}{2}e^{\frac{Fx_{li}}{k_BT}\frac{T}{T_c}\left(1-\frac{T}{T_c}\right)}
			e^{-K_{li}e^{\frac{Fx_{li}}{k_BT}\frac{T}{T_c}}t}	
			+
			\frac{1}{2}e^{\frac{Fx_{li}}{k_BT}\frac{T}{T_c}}
		(K_{li}t)^{\frac{T}{T_c}}\gamma\left(1-\frac{T}{T_c},K_{li}t\right)
		\nn\\
		&
			-
			\frac{1}{2}e^{\frac{Fx_{li}}{k_BT}\frac{T}{T_c}}
		(K_{li}t)^{\frac{T}{T_c}}\gamma\left(1-\frac{T}{T_c},K_{li}e^{\frac{Fx_{li}}{k_BT}\frac{T}{T_c}}t\right)
	+
		\frac{1}{2}e^{-\frac{Fx_{li}}{k_BT}\frac{T}{T_c}}
		\frac{T}{T_c}(K_{li}t)^{-\frac{T}{T_c}}
		\gamma\left(\frac{T}{T_c},K_{li}e^{\frac{Fx_{li}}{k_BT}}t\right)
	\nn\\&\equiv I^{(0)-}_{li}.					
		\label{12102908}
\end{align}
 
Similarly, we obtain 
 \begin{align}
	&
		\int_{-\infty}^\infty d \epsilon_{li}
		\frac{e^{-|\ep_{li}|/k_BT_c}}{2k_BT_c}
		e^{-(\epsilon_{li}-Fx_{li})\Theta(\epsilon_{li}-Fx_{li})/k_BT} 
		e^{
			-K_{li}
		e^{-(\epsilon_{li}-Fx_{li})\Theta(\epsilon_{li}-Fx_{li})/k_BT} t
		}
	\nn\\
	&=
		\frac{1}{2}e^{Fx_{li}/k_BT_c}
	e^{
			-K_{li}t
		}
+
		\frac{1}{2}e^{\frac{Fx_{li}}{k_BT}\frac{T}{T_c}}
		\frac{T}{T_c}(K_{li}t)^{-1+\frac{T}{T_c}}
		\left[
		\gamma\left(1-\frac{T}{T_c},K_{li}t\right)
			-
			\gamma\left(1-\frac{T}{T_c},K_{li}e^{\frac{Fx_{li}}{k_BT}\frac{T}{T_c}}t\right)
		\right]
	\nn\\
	&	
	+
		\frac{1}{2}e^{-\frac{Fx_{li}}{k_BT}\frac{T}{T_c}}
		\frac{T}{T_c}(K_{li}t)^{-1-\frac{T}{T_c}}
		\gamma\left(\frac{T}{T_c}+1,K_{li}e^{\frac{Fx_{li}}{k_BT}}t\right)
	\nn\\&\equiv I^{(1)-}_{li}.					
		\label{12102909}
\end{align}
Substituting Eqs.~(\ref{12102910}), (\ref{11072902}), (\ref{12112605}), (\ref{12112606}), (\ref{12102908}), and (\ref{12102909}) into Eq.~(\ref{wtev2}), we finally obtain the effective waiting time as follows:
\begin{align}
\langle w(t) \rangle_F
&=
	\sum_{k\in \mathcal{N}_+}
	\left\{
	K_{ki}I^{(1)+}_{ki}
	\prod_{j\in \mathcal{N}_+,j\neq k}I^{(0)+}_{ji}
	\prod_{m\in \mathcal{N}_0,m\neq k}I^{(0)0}_{mi}
	\prod_{l\in \mathcal{N}_-,l\neq k}I^{(0)-}_{li}
\right\}
\nn\\
&+
	\sum_{k\in \mathcal{N}_0}
	\left\{
	K_{ki}I^{(1)0}_{ki}	
	\prod_{j\in \mathcal{N}_+,j\neq k}I^{(0)+}_{ji}
	\prod_{m\in \mathcal{N}_0,m\neq k}I^{(0)0}_{mi}
	\prod_{l\in \mathcal{N}_-,l\neq k}I^{(0)-}_{li}
	\right\}
\nn
\\
&+
	\sum_{k\in \mathcal{N}_-}
	\left\{
	K_{ki}I^{(1)-}_{ki}	
	\prod_{j\in \mathcal{N}_+,j\neq k}I^{(0)+}_{ji}
	\prod_{m\in \mathcal{N}_0,m\neq k}I^{(0)0}_{mi}
	\prod_{l\in \mathcal{N}_-,l\neq k}I^{(0)-}_{li}
	\right\}.
\end{align}
Thus, we have derived Eq.~(\ref{11060902}).

	\end{widetext}

	\bibliography{99_refs.bib}	

\end{document}